\documentstyle[pra,aps,amsfonts,epsf]{revtex}

\input{epsf.tex}

\newcommand{\Pint}{\not\hspace{-0.4em}\int}

\newcommand{\s}{\sigma}
\newcommand{\I}{{\rm i}}
\renewcommand{\d}{{\rm d}}
\renewcommand{\a}{\alpha}

\newcommand{\be}{\begin{equation}}
\newcommand{\ee}{\end{equation}}
\newcommand{\bea}{\begin{eqnarray}}
\newcommand{\eea}{\end{eqnarray}}
\newcommand{\ba}{\begin{array}}
\newcommand{\ea}{\end{array}}
\def\J#1#2#3#4{{#1}, {\it #2}, {\bf #3}, #4}
\def\N{Nature}

\def\PAZ{Pis'ma Astron. Zh.}

\def\PRD{Phys. Rev. D}

\def\AJ{Astrophys. J.}
\def\AJL{Astrophys. J. Lett.}

\def\AZ{Astron. Zh.}
\def\MN{M.N.R.A.S.}

\def\ARAA{Ann. Rev. Astron. Astroph.}

\def\CQG{Class. Quantum Grav.}

\begin{document}
\draft
\title{Determining the surface--density distribution\\
in massive galactic disks with a central black hole}

\author{N.~R.~Sibgatullin,\thanks{On leave of absence from the
Department of Hydrodynamics, Lomonosov Moscow State University,
Moscow 119899, Russia} A.~A.~Garc\'\i a, V.~S.~Manko}
\address{Departamento de F\'\i sica, Centro de Investigaci\'on y de
Estudios Avanzados del IPN,\\ A.P.~14-740, 07000 M\'exico D.F.,
Mexico}

\maketitle

\begin{abstract} An efficient method is developed which allows the
calculation of distributions of the surface density in the
equilibrium disk configurations with an isolated point mass at the
center corresponding to known distributions of the angular
velocity. We demonstrate the existence of the upper limit for the
ratio `mass of the galactic disk / mass of the central black hole'
which essentially depends on the form of rotation--curves.
\end{abstract}

\bigskip

\hspace{2cm} {\it Key words:} galaxy disks, rotation curve, black
holes


\section{Introduction}

The super--massive black holes at the nuclei of galaxies and
quasars are able to provide a high luminosity of these objects
thanks to the disk accretion (Lynden--Bell 1969). Nowadays there
exist serious arguments in favor of that hypothesis for a number
of disk--like galaxies (Rees 1998; for the discussion of the
observational data providing evidence about the existence of
compact nuclei in spirals see Rubin and Graham 1987; Sofue 1996;
Ratnam and Salucci 2000), so the problem of finding the
mass--density distributions and total masses of flattened galaxies
possessing a central black hole from the known distributions of
the matter rotation--velocity is of great astrophysical interest.
The rotation--curves are normally constructed from the
measurements of the Doppler effect for the lines $21 cm$, $H_\a$
etc. (see, e.g., Schmidt 1957; de Vaucoulears 1959; Burbidge et
al. 1960; Rubin et al. 1980; Carignan and Freeman 1985; Sofue and
Rubin 2001; Persic and Salucci 1988). The mathematical problem of
reconstruction of the surface mass--density from the
rotation--curves in infinite disks without a central body was
discussed in the papers by Burbidge et al. 1960; Brandt 1960;
Brandt and Belton 1962; Toomre 1963; Binney and Tremaine 1987 in
relation with the problem of non-radiating mass in the disk--like
galaxies. Mestel 1963 proposed models of flattened galaxies as
disks of finite radius with constant angular or linear rotation
velocities.

Another important physical aspect of the problem (we do not touch
it in this paper) is related to the formation of the exterior
massive parts of accretion disks in which the attraction to the
disk because of its mass becomes greater or of the order of the
vertical component of the attraction force from the part of the
black hole. In this case the exterior part of the accretion disk
considerably swells, and the proper field of the disk may play a
substantial role in the thermal balance, distribution of pressure,
etc. (Paczy\'nski 1978; Kozlowski et al. 1979, Kolykhalov and
Sunyaev 1980). For millions years of accretion the mass of the
accretion disk can reach 0.001 the mass of the black hole due to
transition of the angular momentum into the exterior part of the
accretion disk (Kolykhalov and Sunyaev 1980).

In our paper the mass of the accretion disk around a black hole is
neglected, and we only take into account the proper gravitational
field of the galactic disk. In the latter we neglect the influence
of viscosity, pressure and all kind of non--circular movements of
matter, as it is done in the classical articles (Schmidt 1957;
Burbidge et al. 1960; Brandt 1960) because of a small relative
dispersion of velocities. It should be emphasized that the
gravitational field of dark matter and that of the non--disk
component (bulge) exert the same influence on the rotation--curves
as the galactic radiating disk itself. The phenomenological
expressions for these potentials are given in the book of Binney
and Tremaine 1987 and in the paper by Lovelace et al. 1999, but
some researchers are rather sceptic about the methods of
separation of those fields (Burstein and Rubin 1985; Persic and
Salucci 1988).

The aim of our paper is to solve a new Newtonian potential--theory
problem for the gravitational field of a disk extending at some
distance from the central point mass.\footnote{Mention that the
disks with an isolated central mass cannot be obtained as limiting
cases of confocal spheroidal shells of small eccentricity, unlike
the continuous disks previously considered.} The rotation--curve
in the disk is supposed to be known, and it is necessary to find
the corresponding mass of the central object and the distribution
of the surface mass--density in the disk. We give the general
solution of this problem in terms of two successive quadratures
(20), (21) and formulas (19), (23) for the masses of the black
hole and disk, respectively. As an illustration of the application
of general formulas we derive a large class of the asymptotically
Keplerian angular--velocity distributions and demonstrate a strong
dependence of the upper limit of the masses of disks on the choice
of a class of rotation--curves. In spite of a number of strong
assumptions (neglect of a non--disk component, extrapolation of
data about the circular motion of matter into the outer parts of
galactic disks where there is no radiating gas) our formulation
can be considered as a possible approach to the solution of the
problem of a hidden mass in the galactic disks with a black hole
at the center.

\section{The method}

Recall that in the axisymmetric case a thin galactic infinite disk
with an attracting central body of mass $M$ can be described by
the following Newtonian potential (in cylindrical coordinates
$\rho$, $z$)
\be
\phi(\rho,z)=2G\int\limits_0^{\pi}\int\limits_a^\infty
\frac{\s(\rho_0)\rho_0\,\d\rho_0\d\varphi}
{\sqrt{\rho^2+\rho_0^2-2\rho\rho_0\cos\varphi+z^2}}+
\frac{GM}{r}\,, \quad r=\sqrt{\rho^2+z^2}\,, \ee where $G$ is
Newton's gravitational constant. The matter of the disk is
distributed in the plane $z=0$ in the region exterior to the
circumference of radius $a$. From the condition of balance of the
gravitational and centrifugal forces one gets
\be
\omega^2\rho+\left.\frac{\partial\phi}{\partial
\rho}\,\right|_{z\to0} =0 \ee (here we do not consider other
forces like pressure of the gas and the radiation pressure).

Substituting the potential (1) into Eq.~(2), one easily arrives at
the integral equation with the kernel divergent at $\rho=\rho_0$:
\be
G\int\limits_a^\infty\s(\rho_0){\rm K}(\rho,\rho_0)
\d\rho_0=\omega^2\rho-\frac{GM}{\rho^2}\, \ee where
\be
\frac{\rho}{2\rho_0}\,{\rm K}(\rho,\rho_0)= \frac{{\rm
sign}(\rho-\rho_0)E(\tau)}{\rho+\rho_0}+\frac{K(\tau)}{|\rho-\rho_0|}
\,, \quad \tau=-\frac{4\rho_0\rho} {(\rho-\rho_0)^2}\,, \ee
$E(\tau)$ and $K(\tau)$ being the complete elliptic integrals.

Eq.~(3) is a complicated singular integral equation of the
non--Fredholm type. Even the problem of finding $\omega$ for a
given surface density $\s$ would meet difficulties during its
numerical resolution. However, from the point of view of the
observational astronomy, it is likely to solve the inverse
problem, i.e., to determine the mass distribution $\s$ for a known
$\omega$ since from observations it is difficult to establish the
matter distribution in the galactic disk because of the problem of
a hidden (non--radiating) mass,\footnote{In practical approaches
the surface density is normally reconstructed with the aid of the
photometric data (Freeman 1970; for a critical discussion of this
approach see Burstein and Rubin 1985; Persic et al. 1996).}
whereas the rotation--curves are able to provide information about
Keplerian frequencies (Schmidt 1957; Burbidge et al. 1960; Brandt
1960; Binney and Tremaine 1987) thanks to the combined attracting
field. Viewed in this way (finding $\s$ from $\omega$), the
straightforward resolution of Eq.~(3) looks hopeless even by
numerical means.

To circumvent the need of solving the cumbersome Eq.~(3), instead
of (1) we shall take $\phi$ in the form
\be
\phi(\rho,z)=\frac{1}{2\pi^2}
\int\limits_0^{\pi}\d\theta\int\limits_a^\infty
\ln[(s-\rho\cos\theta)^2+z^2]\,\a(s)\,\d s+\frac{GM}{r}\,, \ee
where the real function $\a(s)$ has the meaning of the density of
sources distribution.

This potential satisfies Laplace's equation everywhere except in
the plane $z=0$, the sources of the disk being distributed outside
the circle $\rho<a$. The main advantage of such a representation
over (1) is that it allows to introduce into the problem under
consideration the powerful theory of analytic functions in the
complex plane since the first term in (5) can be cast into the
form $\int_0^\pi f(z+\I\rho\cos\theta)\,\d\theta$.

Let us show that
\be
\left.\frac{\partial\phi}{\partial z}\right|_{z\to+0}= \left\{
\begin{array}{ll}0\,,&0<\rho<a\\
\frac{1}{\pi}\int\limits_a^\rho\frac{\a(s_0)\d s_0}
{\sqrt{\rho^2-s_0^2}}\,,&a\le\rho\,.\\ \end{array}\right. \ee

Indeed, from (5) we have
\be
\frac{\partial\phi}{\partial z}=\frac{1}{2\pi^2}
\int\limits_0^{\pi}\d\theta\int\limits_a^\infty\left(
\frac{-\I}{s-\rho\cos\theta-\I z}+\frac{\I}{s-\rho\cos\theta+\I
z}\right) \a(s)\,\d s-\frac{GMz}{r^3}\,. \ee

Tending $z\to +0$ for $\rho\cos\theta=s_0>a$ and using the
Sokhotsky--Plemelj formula for the Cauchy-type integrals,
namely\footnote{Here $s_0$ is a point on a smooth contour ${\cal
L}$: $s_0\in{\cal L}$; $Z\to+s_0$ means that the point $Z$ is
approaching the point $s_0$ from the left side in the sense of a
positive running of the curve ${\cal L}$; the symbol
$\int\hspace{-0.38cm}-$ denotes the principal value of the
respective integral.}
\be
\lim_{Z\to s_0}\int\limits_{\cal L} \frac{\a(s)\d s}{s-Z}=
\Pint\limits_{\cal L} \frac{\a(s)\d s}{s-Z}\pm\pi \I\a(s_0), \ee
we get (in our case $Z=s_0\pm\I z$)
\be
\left.\frac{\partial\phi}{\partial z}\right|_{z\to+0}= \left\{
\begin{array}{ll}0\,,&0<s_0<a\\
\frac{1}{\pi}\int\limits_0^\pi \a(s_0)\d\theta \,,&s_0\ge a\,.\\
\end{array}\right. \ee

From (9) the formula (6) follows immediately.

Let us consider the dependence of $\phi$ on $\rho$ when $\rho>a$
and $z\to 0$. From (8) we obtain
\be
\left.\frac{\partial\phi}{\partial\rho}\right|_{z\to 0}=
-\frac{GM}{\rho^2}+\frac{1}{\pi\rho}\left(\int\limits_a^\infty
\a(s)\d s-\frac{1}{\pi}\int\limits_a^\infty\a(s)s\,\d
s\Pint\limits_{-\rho}^\rho \frac{\d
\rho'}{(s-\rho')\sqrt{\rho^2-\rho'^2}}\right)\,. \ee Using now the
formula
\be
\frac1\pi\Pint\limits_{-\rho}^\rho \frac{\d
\rho'}{(s-\rho')\sqrt{\rho^2-\rho'^2}}= \left\{
\begin{array}{ll}0\,,&a<s<\rho\\
\frac{1}{\sqrt{s^2-\rho^2}}\,,&s>\rho\,,\\
\end{array}\right. \ee Eq.~(10) can be rewritten in the
form
\be
\left.\frac{\partial\phi}{\partial\rho}\right|_{z\to 0}=
-\frac{GM}{\rho^2}+\frac{1}{\pi\rho}\left(\int\limits_a^\infty
\a(s)\d s-\int\limits_\rho^\infty \frac{\a(s)\,s\,\d s}
{\sqrt{s^2-\rho^2}}\right)\,. \ee

Let us show that for a finite mass of the disk, the integral
$\int_a^\infty \a(s)\d s$ should vanish.

Indeed, from (6) we have for $\rho\ge a$
\be
-4\pi G \s(\rho)=\left[\frac{\partial\phi}{\partial z}\right]
=2\left.\frac{\partial\phi}{\partial z}\right|_{z\to+0} =
\frac2\pi\int\limits_a^\rho\frac{\a(s_0)\d s_0}
{\sqrt{\rho^2-s_0^2}}\,,\ee hence the mass of the disk is equal to
\be
M_d=-\frac{1}{4\pi
G}\int\limits_0^{2\pi}\d\varphi\int\limits_a^\infty
\left[\frac{\partial\phi}{\partial z}\right]\rho\,\d\rho
=-\frac{1}{G}\int\limits_a^\infty m(\rho)\,\d\rho, \quad
m(\rho)=\frac{1}{\pi}\int\limits_a^\rho \frac{\a(s_0)\rho\,\d s_0}
{\sqrt{\rho^2-s_0^2}}\,.\ee

In order the mass of the disk be finite, it is necessary that
\be
\lim_{\rho\to\infty}m(\rho)=0 \quad \Rightarrow \quad
\int\limits_a^\infty \a(s)\,\d s=0\,. \ee

Substituting now (12) into the equilibrium condition (2), we
obtain
\be
\omega^2\rho-\frac{GM}{\rho^2}=\frac{1}{\pi\rho}
\int\limits_\rho^\infty \frac{\a(s)s\,\d s}{\sqrt{s^2-\rho^2}}\,.
\ee

After introducing the new variables $t=a^2/s^2$, $x=a^2/\rho^2$,
the last equation reduces to the Abel integral equation
\be
\frac{a^2}{2\pi}\int\limits_0^x\frac{\a(t)\,\d t}
{t^{3/2}\sqrt{x-t}} =\frac{a^3\omega^2}{x^{3/2}}-GM\,, \ee and the
well--known formula for the solution of this equation yields (see,
e.g., Sneddon 1956, p.~318)
\be
\frac{a^2\a(x)}{2x^{3/2}}=\frac{\d}{\d x}\int\limits_0^x \frac{\d
t}{\sqrt{x-t}}\left(\frac{a^3\omega^2}{t^{3/2}}-GM\right)\,. \ee

The substitution of (18) into the condition (15) yields the
formula for the mass of the central body
\be
M=\frac{a^3}{2G}\int\limits_0^{1}\frac{\omega^2\,\d
t}{t^{3/2}\sqrt{1-t}}\,, \ee hence the function $\a(x)$ can be
expressed through only the known distribution of the angular
velocity:
\be
\frac{\a(x)}{2ax^{3/2}}=\frac{\d}{\d x}\int\limits_0^x
\frac{\omega^2\,\d
t}{t^{3/2}\sqrt{x-t}}-\frac{1}{2\sqrt{x}}\int\limits_0^{1}
\frac{\omega^2\,\d t}{t^{3/2}\sqrt{1-t}}\,. \ee

The corresponding distribution of the surface density, obtainable
from (13), assumes the form
\be
-4\pi G\s(x)=\frac{\sqrt{x}}{\pi}\int\limits_x^{1} \frac{\a(t)\,\d
t}{t\sqrt{t-x}}\,. \ee

From (21), (20) follows that the asymptotics of the surface
density near the inner rim of the disk $\rho=a$ is given by the
formula
\be
-4\pi^2 G\s(x)\approx A\sqrt{1-x}\,, \quad A=a \int\limits_0^1
\left[\left(\frac{\omega^2}{t^{3/2}}\right)_{,\,t}
-\frac{1}{2}\left(\frac{\omega^2}{t^{3/2}}
-\frac{G(M+M_d)}{a^3}\right)\right]\frac{\d t}{\sqrt{1-t}}\,.\ee

It can be seen from (22) that in the presence of a central body
possessing a finite mass the surface density near the inner rim of
the disk has no finite limit when $a$ tends to zero. The
inequality $A\le 0$ imposes a physical restriction on the
angular--velocity distribution.

Therefore, we have finally succeeded in expressing the
mass--density distribution in the disk exclusively through the
supposedly known distribution of the angular velocity via two
quadratures which can be easily taken numerically in the general
case, and analytically for large classes of particular
distributions.

The total mass of the disk, $M_d=2\pi\int\limits_a^{\infty}\s
\rho\,\d\rho$, is obtainable from (18) and (21):
\be
GM_d=\lim_{t\to0}\frac{a^2\a(t)}{2t}=
\lim_{t\to0}\frac{a^3\omega^2}{t^{3/2}}-GM\,. \ee

To give an analytical illustration of the results obtained, let us
consider the following smooth distribution of the angular velocity
possessing the Keplerian asymptotics:\footnote{In (24) the
coefficient $a_1$ is equal to zero by virtue of equations (17) and
(18).}
\be
\omega^2\rho^3=a_0+a_2\frac{a^2}{\rho^2}
+a_3\frac{a^3}{\rho^3}+\ldots+a_n\frac{a^{n}}{\rho^n} \equiv
a_0+\sum\limits_{k=2}^n a_k t^{k/2}\,, \ee where $a_k$ are some
constant coefficients.

Substituting this expression into (18) we arrive at
\be
\frac{\a(x)}{x^{3/2}} = \frac{2}{\sqrt{x}}\,(a_0-GM)
+\sum\limits_{k=2}^n(k+1)
x^{(k-1)/2}B\Bigl(\frac12,1+\frac{k}{2}\Bigr)a_k\,, \ee $B(x,y)$
denoting the Euler beta--function.

From the condition (19) we obtain the mass of the black hole:
\be
2GM = 2a_0 +\sum\limits_{k=2}^n
B\Bigl(\frac12,1+\frac{k}{2}\Bigr)a_k\,. \ee

Substituting the expression (25) into formula (21) one finds
\be
-4\pi^2 G\frac{\s(x)}{\sqrt{x}}=4(a_0-GM)\sqrt{1-x}+
\sum\limits_{k=2}^n(k+1)B\Bigl(\frac12,1+\frac{k}{2}\Bigr)a_kJ_k\,,
\quad J_k\equiv\int\limits_x^1\frac{t^{k/2}\d t}{\sqrt{t-x}}\,.
\ee

Using integration by parts, for the integrals $J_k$ it is easy to
obtain the recurrent formula
\be
(k+1)J_k=2\sqrt{1-x}+kxJ_{k-2}\,, \ee the use of which in (27)
gives the explicit expression for the surface density: \bea
\s(x)&=&-\frac{x^{3/2}\sqrt{1-x}}{2\pi^2G}\sum\limits_{k=2}^n
B\Bigl(\frac12,1+\frac{k}{2}\Bigr)a_k\left\{\frac{k}{k-1}
+x\,\frac{k(k-2)}{(k-1)(k-3)}+\ldots\right. +x^{[k/2]-1}
\nonumber\\ &\times&\frac{k(k-2)\ldots(k-2[k/2]+2)}
{(k-1)(k-3)(k-2[k/2]+1)} \left.\left(
1+\frac{(k-2[k/2])x}{\sqrt{1-x}}\,
\ln\frac{1+\sqrt{1-x}}{\sqrt{x}}\right)\right\}\,. \eea

The inequality $A\le0$ (cf. (22)) imposes the following
restriction on the parameters $a_k$:
\be
\sum\limits_{k=2}^n a_k\left[\frac{k}{2}B\left(\frac{1}{2},
\frac{k}{2}\right)-\frac{1}{2}B\left(\frac{1}{2},
1+\frac{k}{2}\right)\right]\le 0\,.\ee

Mention that when the function $\omega^2\rho^3$ is representable
by a polynomial in inverse powers of $\rho^2$, $a_{2k+1}=0,
k=1,2,...,N$, then the surface density is given by the formula
\bea \s(x)&=&\frac{x^{3/2}\sqrt{1-x}}{\pi^2a^2G}
\sum\limits_{s=1}^N\beta_s x^{s-1}\,,\nonumber \\
\beta_s&=&\sum\limits_{k=s}^N\sum\limits_{m=0}^s
\frac{2^k(k!)^2(-1)^{m+1}
a_{2k}}{(2k-1)!!m!(k-s)!(s-m)!(2k+2m-2s+1)}\,. \eea

\section{Some examples}

(A) Let us consider the case when the rotation--curves belong to
the family of curves with Keplerian asymptotics of the type
\be
\omega^2\rho^3=GM+GM_d(1+b_1x+b_2x^2+b_3x^3)\,, \quad
a_{2i}=GM_db_{i}\,, \quad x=a^2/\rho^2\,. \ee

From the formula (26) follows that
\be
b_1=-\frac{3}{2}-\frac{4}{5}\,b_2-\frac{24}{35}\,b_3\,. \ee

Taking into account (33), one obtains from (31) the distribution
of the surface density corresponding to (30):
\be
\s(\rho)=\frac{aM_d\sqrt{1-a^2/\rho^2}}{\rho^3}
\left[2+\frac{16}{45}b_2+\frac{64}{175}b_3-
\left(\frac{64}{45}b_2+\frac{128}{175}b_3\right)x
-\frac{256}{175}b_3x^2\right]\,. \ee

Formulas (32)--(34) fully describe the two--parameter family of
the angular--velocity distribution and the corresponding
distribution of the surface density. In Fig.~1 we have plotted the
region $D$ of the plane $(b_2,b_3)$ on which the right--hand side
of (34) is a positive quantity, so that (32) has physical sense.
The boundary of the region $D$ consists $(i)$ of two line segments
$AB$ and $AC$ tangent to the ellipse at the points $B$ and $C$,
their defining equations are
$1+\frac{8}{45}b_2+\frac{32}{175}b_3=0$ and
$1-\frac{8}{15}b_2-\frac{32}{35}b_3=0$;\footnote{The inequality
$A\le 0$ is equivalent to the condition
$1-\frac{8}{15}b_2-\frac{32}{35}b_3\ge0$.} $(ii)$ of the part of
the ellipse between the points $B$ and $C$; the equation of the
ellipse results by setting to zero the discriminant of the
quadratic polynomial at the right--hand side of (34).

Let us call $\mu(b_2,b_3)$ the minimal value of the cubic
polynomial $f(x)$:
\be
f(x)\equiv 1-\left(\frac32+\frac{4}{5}b_2+
\frac{24}{35}b_3\right)x+b_2x^2+b_3x^3 \ee on the interval
$(0,1]$. Then the maximally possible mass of the disk can be
calculated from the condition of the non--negativeness of the
right--hand side of (32):
\be
M_{d\,max}(b_2,b_3)=-M/\mu(b_2,b_3)\,. \ee

The polynomial $f(x)$ can assume its minimal value either inside
of the interval $(0,1)$ or at the boundary $x=1$. In Fig.~1 the
dashed line separates the region $D$ into two parts $D_1$ and
$D_2$. Inside of $D_1$, where $f(1)$ is the minimal value of
$f(x)$ on the interval $(0,1)$, the function $\mu(b_2,b_3)$ has
the simple form $\mu=-\frac12+\frac15 b_2+\frac{11}{35}b_3$.

In Fig.~2 the curves $M_{d\,max}$ are shown as functions of $b_2$
for different fixed values of $b_3$, the points $(b_2,b_3)$
belonging to $D$. A simple investigation shows that the maximally
possible mass of the disk can reach $9M$. It is interesting to
point out that the more complicated structure has the
angular--velocity distribution in the disk (i.e. the greater is
$n$ in the polynomial (31)), the larger mass the disk can have in
principle: for the one--parameter family of the angular--velocity
distribution the maximal mass of the disk is equal to $5M$, and
for the two--parameter family the maximal value elevates up to
$9M$.

Let us consider three particular cases of the distribution (32).

1) The simplest particular case, $b_2=b_3=0$, was found by Lemos
and Letelier 1994\footnote{See the discussion of the corresponding
gravitational potential in the paper by Semer\'ak and
\u{Z}a\u{c}ek 2000.}. In this case
\be
\omega^2\rho^3=GM+GM_d\Bigl(1-\frac32
\frac{a^2}{\rho^2}\Bigr), \quad
\s(\rho)=\frac{2aM_d\sqrt{\rho^2-a^2}}{\pi^2\rho^4}\,. \ee

From (35) and (36) follows that the mass of the disk cannot exceed
two masses of the central body.

2) The particular case $b_2=15/8$, $b_3=0$ corresponds to the
distribution of the angular velocity and surface density of the
form
\be
\omega^2\rho^3=GM+GM_d\Bigl(1-3
\frac{a^2}{\rho^2}+\frac{15}{8}\frac{a^4}{\rho^4}\Bigr)\,, \quad
\s(\rho)=\frac{8aM_d(\rho^2-a^2)^{3/2}}{3\pi^2\rho^6}\,, \ee and
the mass of the disk does not exceed $5M$.

3) The choice $b_2=45/8$, $b_3=-35/16$ gives
\be
\omega^2\rho^3=GM+GM_d\Bigl(1-\frac{9}{2}
\frac{a^2}{\rho^2}+\frac{45}{8}\frac{a^4}{\rho^4}-
\frac{35}{16}\frac{a^6}{\rho^6}\Bigr)\,, \quad
\s(\rho)=\frac{16aM_d}{5\pi^2\rho^3}
\left(1-\frac{a^2}{\rho^2}\right)^{5/2}\,, \ee and the mass of the
disk cannot be greater than $6.69M$.

(B) To demonstrate that the mass of the galactic disk can in
principle be arbitrarily large compared to the mass of the central
black hole, let us consider the following
rotation--curves\footnote{The rotation--curves (40) were first
considered by Kuzmin 1956; they enter as a particular case into
the class of curves possessing the Keplerian asymptotics at
$\rho\to\infty$ proposed by Brandt 1960 for infinite disks without
a black hole at the center.}
\be
\omega^2(\rho^2+c^2)^{3/2}=G(M+M_d)\,, \quad c={\rm const}. \ee

In this case the equation (23) is fulfilled automatically. Let us
denote the parameter $c^2/a^2$ as $b$.

Calculating $\a(x)$ according to formula (20), we get
\be
\frac{\a(x)}{2ax^{3/2}}=-\frac{b(-1+(3+b)+bx^2)}
{\sqrt{x}(1+b)(bx+1)^2}\,\frac{G(M+M_d)}{a^3}\,. \ee

Substituting the expression (41) into (21), we obtain the formula
for the distribution of the surface density:
\be
\s(x)=\frac{(M+M_d)\sqrt{bx^3}}{\pi^2 a^2}\left(
\frac{\sqrt{b(1-x)}}{(1+b)(1+bx)} +\frac{{\rm
Arctan\sqrt{b(1-x)/(1+bx)}}}{(1+bx)^{3/2}}\right)\,. \ee

Then from (19) follows a surprisingly simple formula for the mass
of the disk:
\be
M_d=bM\,. \ee

The parameter $b$ in formulas (41)--(43) can be an arbitrary
positive number. Therefore, the mass of the galactic disk can be
by far greater than the mass of the black hole (if the linear
velocity $V=\rho\omega$ achieves its maximum at the distances much
larger than the accretion part of the disk $b\gg 1$). Mention that
when $a\to 0$, the variable $x$ tends to zero too, while the
parameter $b$ tends to infinity. From (43) then follows that $M\to
0$, and from formula (42) we obtain Kuzmin--Toomre's result for
the surface--density distribution (40):
\be
\s(\rho)=\frac{c\,M_d}{2\pi(\rho^2+c^2)^{3/2}}\,. \ee

\section{Conclusion}

Therefore, we have succeeded in solving the problem of the
determination of the surface density in massive galactic disks
with a central massive black hole from the known rotation--curves.
The problem is reduced to finding two successive quadratures (20)
and (21) which can be easily calculated by numerical means in the
most general case of smooth distributions of the angular velocity.
Besides, we have shown how the knowledge of a rotation--curve
permits to calculate the mass of the central body and that of the
galactic disk.

The analysis of the formulas obtained gives rise to the following
two general observations.

\medskip

($i$) The mass of the disk around a central body in each case
should be less than some upper limiting value which depends on the
degree of concentration of matter at the inner boundary of the
disk. The mass of the disk strongly depends on the form of the
rotation--curve and in principle can exceed the mass of the black
holes many times.

\medskip

($ii$) For an arbitrary non--degenerated smooth distribution of
the angular velocity in the disk, the surface density at the inner
boundary tends to zero like $\sqrt{1-a^2\rho^{-2}}$. It
corresponds to the parabolic profile of thickness of the disk for
a finite value of the volume density at the inner boundary of the
disk. At infinity the surface density decreases as $\rho^{-3}$.

\medskip

Lastly, it should be remarked that the method developed in this
paper is not restricted to only the case of infinite massive disks
with attractive central body, but also permits to treat the cases
of finite disks and systems of rings. These last cases are in the
stage of analysis and preparation now, and will become a subject
of future publications.

\section*{Acknowledgments}

We are thankful to the anonymous referee and to Prof.~A.~Toomre
for valuable comments and suggestions. This work has been
partially supported by Projects 34222--E and 38495--E from CONACyT
of Mexico. N.S. acknowledges financial support from SNI--CONACyT,
Mexico.

\begin{figure}[htb]
\centerline{\epsfysize=13cm\epsffile{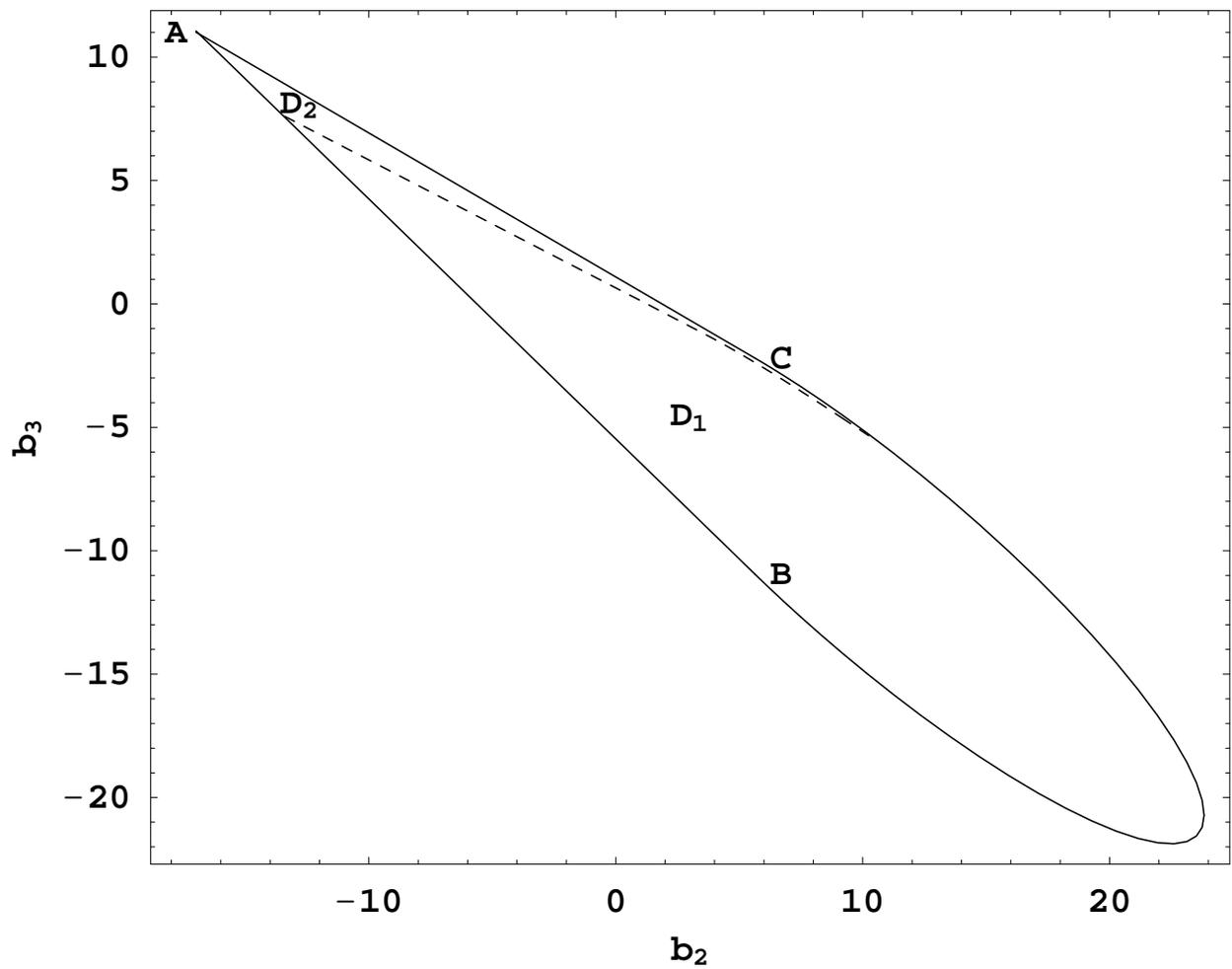}} \caption{The
region of positiveness of $\s(\rho)$ in the plane $(b_2, b_3)$.}
\end{figure}

\begin{figure}[htb]
\centerline{\epsfysize=11cm\epsffile{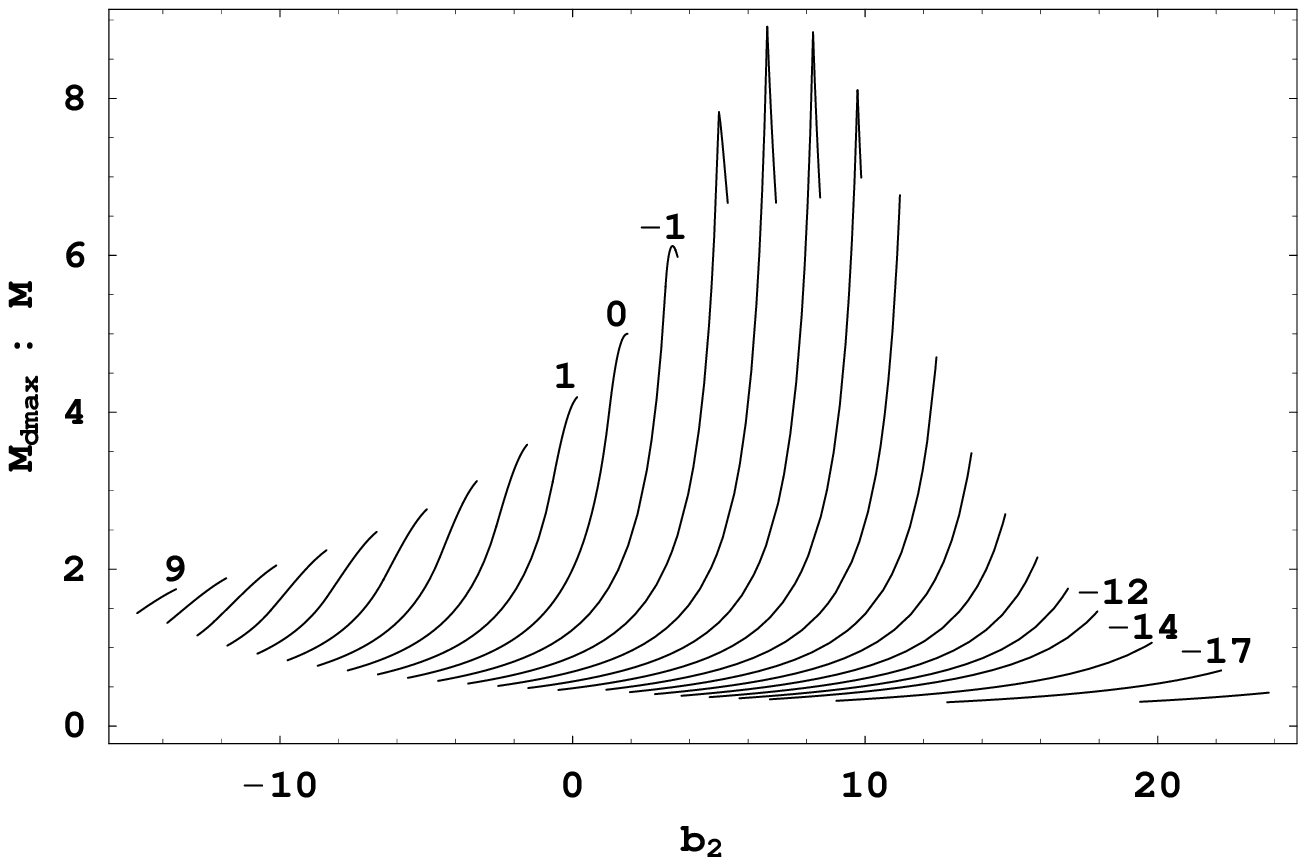}} \caption{The
dependence of $M_{d max}/M$ on $b_2$ for different fixed values of
$b_3$ $(b_3=-21,-17,-14,-12,...,-1,0,1,...,9)$.}
\end{figure}


\begin{references}

\item[] {\it J. Binney, S. Tremaine},
Galactic Dynamics, Princeton: Princeton Univ. Press, 1987, p. 42.

\item[] {\it J.C. Brandt}, \AJ, {\bf 131}, 293
(1960).

\item[] {\it J.C. Brandt, M.J.S. Belton},
\AJ, {\bf 136}, 352 (1962).

\item[] {\it E.M. Burbidge, G.R. Burbidge, K.H. Prendergast}, \AJ, {\bf
131}, 282 (1960).

\item[] {\it D. Burstein, V.C. Rubin}, \AJ, {\bf
297}, 423 (1985).

\item[] {\it C. Carignan, K.C. Freeman},
\AJ, {\bf 294}, 494 (1985).

\item[] {\it K.C. Freeman},
\AJ, {\bf 160}, 811 (1970).

\item[] {\it P.I. Kolykhalov, R.A.
Sunyaev}, \PAZ, {\bf 6}, 680 (1980).

\item[] {\it M. Kozlowski, P.J. Wiita, B. Paczy\'nski}, Acta Astron., {\bf
29}, 157 (1979).

\item[] {\it G.G. Kuzmin}, \AZ, {\bf 33}, 27
(1956).

\item[] {\it J.P.S. Lemos, P.S. Letelier},
\PRD, {\bf 49}, 5135 (1994).

\item[] {\it R.V.E. Lovelace, L. Zhang, D.A.
Kornreich, M.P. Haynes}, \AJ, {\bf 524}, 634 (1999).

\item[] {\it D. Lynden--Bell}, \N, {\bf 223},
690 (1969).

\item[] {\it L. Mestel}, \MN, {\bf 126}, 553
(1963).

\item[] {\it B. Paczy\'nski}, Acta Astron., {\bf
28}, 91 (1978).

\item[] {\it M. Persic, P. Salucci}, \MN,
{\bf 234}, 131 (1988).

\item[] {\it M. Persic, P. Salucci, F. Stel},
\MN, {\bf 281}, 27 (1996).

\item[] {\it Ch. Ratnam, P. Salucci},
astro-ph/0008121.

\item[] {\it M.J. Rees}, Black Holes and Relativistic
Stars, Chicago: University of Chicago Press, 1998, p. 79.

\item[] {\it V.C. Rubin, W.K. Ford, N.
Thonnard}, \AJ, {\bf 238}, 471 (1980).

\item[] {\it V.C. Rubin, J.A. Graham}, \AJL,
{\bf 316}, 67 (1987).

\item[] {\it M. Schmidt}, BAN, {\bf 14}, 17 (1957).

\item[] {\it O. Semer\'ak, M. \u{Z}a\u{c}ek},
\CQG, {\bf 17}, 1613 (2000).

\item[] {\it I.N. Sneddon}, The Use of Integral
Transforms, New York: McGraw--Hill Book Company, 1956.

\item[] {\it Y. Sofue}, \AJ, {\bf 458}, 120 (1996).

\item[] {\it Y. Sofue, V.C. Rubin}, \ARAA, {\bf
39}, 137 (2001).

\item[] {\it A. Toomre}, \AJ, {\bf 138}, 385 (1963).

\item[] {\it G. de Vaucouleurs}, Handbuch der
Physik, {\bf 53}, 348 (1959).

\end{references}
\end{document}